\documentclass[preprintnumbers,amsmath,amssymbm,prd]{revtex4}
\usepackage{epsfig}
\usepackage{graphicx}
\usepackage{amssymb}

\begin{document}

\title{Bekenstein, I, and the quantum of black-hole surface area}
\author{Shahar Hod}
\affiliation{The Ruppin Academic Center, Emeq Hefer 40250, Israel}
\affiliation{ } \affiliation{The Hadassah Institute, Jerusalem
91010, Israel}
\date{\today}
{\it In memory of Prof. Jacob Bekenstein, whose insightful and
elegant works have inspired my research}

\begin{abstract}

\ \ \ Professor Jacob Bekenstein was known not only for his
brilliant and original physical ideas, but also for their clear
presentation in his lectures and seminal research papers. I here
provide a short review of Bekenstein's pioneering ideas about the
quantization of black holes. I also describe my attempt, as a young
and extremely naive student, to prove him wrong and how I got
convinced in the correctness and utility of his deep physical
intuition. Finally, my personal contribution to the ongoing attempts
to understand the evenly spaced (discrete) area spectrum of
quantized black holes, as originally suggested by Bekenstein in the
early days of his scientific career, is described.
\end{abstract}
\bigskip
\maketitle

\section{Preface}

The remarkably deep and unique physical insights of Professor Jacob
Bekenstein, that were also characterized by extremely beautiful
mathematical elegancy, have inspired the imagination of many
physicists, I among them, during the last decades. The sudden death
of Jacob has left a huge hole in many hearts, but his original ideas
still live with us. I am glad and honored for the invitation to
contribute a personal essay for the special memorial volume for
Jacob Bekenstein.

\section{The Bekenstein quantization of the black-hole surface area}

The quantization of the black-hole horizon area was first discussed
by Jacob Bekenstein, then a young and brilliant researcher in
Princeton university \cite{NoteAus}, more than four decades ago
\cite{Bek1}. Bekenstein has based his original and ground
breaking quantization argument on the
intriguing physical fact discussed in \cite{Chri} that, while the
conserved charges of a Kerr-Newman black hole (mass, electric
charge, and angular momentum) may change due to the absorption of
test particles, the surface area of the black hole may remain
unchanged during an assimilation process of a point (structureless)
particle if the later is absorbed at the black-hole horizon while being in a
turning point \cite{Notetp} of its radial motion.

It was therefore suggested by Bekenstein that the black-hole surface
area may serve as a classical adiabatic invariant. In particular, in
the spirit of the Ehrenfest principle that has been formulated in
the early days of quantum mechanics \cite{Born}, Bekenstein has
boldly conjectured that the surface area of a black hole should have
a discrete (quantized) spectrum of the form \cite{Bek1,Noteunit}
\begin{equation}\label{Eq1}
A_n=\gamma\hbar\cdot n\ \ \ ; \ \ \ n=1,2,3,...\  ,
\end{equation}
where $\gamma$ is a dimensionless constant.

Bekenstein has further suggested \cite{Bek1} to estimate the spacing
of the black-hole area levels [that is, the value of the
dimensionless physical coefficient $\gamma$ in (\ref{Eq1})] from a
semi-classical gedanken experiment in which a black hole absorbs a finite-size
neutral particle whose radial location and momentum are related by
the quantum Heisenberg uncertainty principle \cite{Bek2,NoteBek2}.
In particular, Bekenstein has explicitly proved that, taking
cognizance of quantum effects that influence the motion of test
particles in black-hole spacetimes \cite{Bek1,Bek2,NoteBek2}, the
minimal increase in the black-hole horizon area is given by the
simple expression
\begin{equation}\label{Eq2}
(\Delta A)_{\text{min}}=8\pi\hbar\  .
\end{equation}

Intriguingly, the minimal increase (\ref{Eq2}) in the black-hole
surface area, as first suggested by Bekenstein in his seminal
works \cite{Bek1,Bek2}, is {\it universal}
in the sense that it does not depend on the physical parameters of
the black hole and the absorbed particle that have been used to
derive it. This physically interesting observation naturally
suggests that a lower bound of the form (\ref{Eq2}) is of
fundamental importance in a quantum theory of gravity \cite{Bek1,Bek2}.

\section{Challenging (and then confirming) the Bekenstein area quantization}

In 1997, while I was investigating the inner structure of black
holes under the insightful and enjoyable supervision of Professor
Tsvi Piran at the Hebrew university of Jerusalem, I had a casual
conversation with Dr. Avraham Mayo, then a student of Professor
Bekenstein. The short and (for me) life changing conversation took
place at the corridor of the physics department, few steps from
Jacob's office. Until that moment I only knew Professor Bekenstein
by his name. In particular, at that time I had no real information
about Bekenstein's research interests nor about his incredible
scientific achievements. Mayo had told me that his supervisor,
Professor Bekenstein, is working on the quantization of black holes
\cite{NoteAvi}.

The intriguing concept of bringing the ideas of quantum physics
into the realms of black-hole physics and gravitation was totally
new for me and it had immediately caught my imagination.
I remember myself running, after the short conversation with Mayo,
to the library in order to search for Bekenstein's works
on the quantization of black holes \cite{Bek1}. I have read his
seminal papers with much excitement and passion. The papers were
written in a remarkably clear style (which, as I later learned,
characterizes the entire set of Jacob's scientific works). In particular,
the brilliant and (for me) new physical ideas were presented
in these seminal papers in a mathematically elegant fashion.

In the first few days after reading, for the first time,
Bekenstein's papers on the quantization of the black-hole horizon
area, I was  incredibly enthusiastic and happy about the beauty of
physics that was reflected to me from his papers. In particular,
the remarkably simple and mathematically
elegant quantum area spectrum (\ref{Eq1}) that emerged from
Bekenstein's deep physical insights \cite{Bek1} seemed to me as a
new and exciting window into the elusive world of quantum gravity.

However, as a young, bold, and probably extremely naive student, I
took it as a scientific, and even more, as a personal challenge to prove that the physically
intriguing conclusion of Prof. Bekenstein, regarding the discrete
quantization of the black-hole surface area, is wrong. In particular,
I have noticed that the interesting analysis of Bekenstein presented
in his seminal work \cite{Bek1} is restricted to the regime of {\it
neutral} particles which are absorbed by black holes. I had the
feeling that an analogous gedanken experiment, which involves the
absorption of {\it charged} (rather than neutral) particles by
charged black holes, may provide a physically interesting
counter-example to the intriguing lower bound (\ref{Eq2}) derived by
Bekenstein on the increase in black-hole surface area.

Analyzing the absorption of a charged particle of proper mass $\mu$
and electric charge $q$ by a charged Reissner-Nordstr\"om black hole
of mass $M$ and electric charge $Q$, I have found that the minimally
allowed increase in the black-hole surface area is given
by the simple expression \cite{Hod1}
\begin{equation}\label{Eq3}
(\Delta A)_{\text{min}}={{4\pi\mu^2}\over{qE_+}}\  ,
\end{equation}
where $E_+=Q/r^2_+$ is the electric field of the charged black hole
at its surface $r_+=M+(M^2-Q^2)^{1/2}$.

Intriguingly, at first glance one may deduce from the analytically
derived expression (\ref{Eq3}) that, for charged black holes, the
minimal increase $(\Delta A)_{\text{min}}$ in surface area may
become arbitrarily small in the $E_+\to\infty$ limit. For two days I
was very excited from the (obviously naive) thought that I, a young
researcher, have found a counter-example to Bekenstein's well known
lower bound (\ref{Eq2}).

However, I have then realized that, within the framework of a quantum
theory of gravity, the black-hole electric field cannot become
arbitrarily large. In particular, vacuum polarization effects
associated with the Schwinger pair production mechanism \cite{Schw}
set the upper bound \cite{EL}
\begin{equation}\label{Eq4}
E_+\leq {{\pi\mu^2}\over{q\hbar}}\
\end{equation}
on the strength of the black-hole electric field. Substituting
(\ref{Eq4}) into (\ref{Eq3}), I have found the simple lower bound
\cite{Hod1}
\begin{equation}\label{Eq5}
(\Delta A)_{\text{min}}=4\hbar\
\end{equation}
on the increase in surface area of charged black holes due to the
absorption of charged particles.

The fact that the lower bound (\ref{Eq5}), which I have derived for
the minimal increase in the horizon area of charged black holes, and
the lower bound (\ref{Eq2}) which, as originally proved by Bekenstein \cite{Bek1},
provides the minimal increase in black-hole surface area due to the
absorption of neutral particles, are of the same
order of magnitude has convinced me that there is something very
deep and physically correct in Bekenstein's intriguing ideas about
the quantization of black holes.

\section{The Bohr correspondence principle and black-hole area
quantization}

The intriguing fact that both expressions (\ref{Eq2}) and
(\ref{Eq5}) for the minimal increase in the black-hole surface area
are universal (that is, independent of the physical parameters of
the black hole and the absorbed particle) clearly indicates that
these lower bounds may be of fundamental physical importance in
black-hole physics and in the elusive quantum theory of gravity.

It should be realized, however, that the precise value of the
fundamental dimensionless constant $\gamma$ in the quantized
Bekenstein area spectrum (\ref{Eq1}) cannot be determined by the
gedanken experiments of \cite{Bek1,Hod1}. In particular, due to the
inherent fuzziness of the Heisenberg uncertainty principle and due
to the approximated nature of the expression (\ref{Eq4}) for the
critical electric field, the estimated lower bounds (\ref{Eq2}) and
(\ref{Eq5}) on the black-hole area increase can be challenged.

Furthermore, Bekenstein and Mukhanov \cite{Muk,BekMuk} have
correctly argued that the thermodynamic Bekenstein-Hawking area-entropy relation \cite{Bek1,Haw}
\begin{equation}\label{Eq6}
S_{\text{BH}}={{A}\over{4\hbar}}\
\end{equation}
for black holes, together with the Boltzmann-Einstein formula
\begin{equation}\label{Eq7}
g_n=e^S\in\mathbb{N}
\end{equation}
for the number of microstates in statistical physics, imply that the
value of the dimensionless physical constant $\gamma$ in the
quantum spectrum (\ref{Eq1}) should be of the simple functional form
\begin{equation}\label{Eq8}
\gamma=4\ln k\ \ \ \ \text{with}\ \ \ \ k=2,3,...\  .
\end{equation}

Although physically convincing, the combined
thermodynamic-statistical-physics argument presented in
\cite{Muk,BekMuk} could not fix uniquely the value of the integer
$k$ in the functional relation (\ref{Eq8}). As a researcher taking
his first steps in the exciting academic world, I felt that it is a
physically worthy challenge to try to determine the value of this
mysterious integer.

In particular, one of the physical questions that most intrigued me
as a young student was how the observed (classical) world emerges
from the fundamental (quantum) description of nature. At that time,
I already knew that, using the physical concept of coherent states,
one can learn about the classical description of macroscopic
phenomena from the underlying microscopic quantum description. But
can one learn about the fundamental quantum world from its
macroscopic coarse-grained classical description?

It took me a few days to realize that, luckily enough, the answer to
the above stated question is `Yes!'. In particular, the Bohr
correspondence principle that was formulated in the early days of
quantum mechanics \cite{Born} allows one to relate, in the regime of
large quantum numbers, the fundamental discrete properties of a
quantum physical system to the corresponding macroscopic (classical)
properties of the system.

In particular, I have realized that the classical resonant
oscillation frequencies of black-hole spacetimes can be related,
through the Bohr correspondence principle, to the underlying quantum
properties of these fundamental physical objects \cite{Hod2,Hodbr}.
Taking cognizance of the Bekenstein quantized area spectrum
(\ref{Eq1}) and the area-mass relation $A=16\pi M^2$ for
Schwarzschild black holes \cite{Bek1}, one finds the characteristic
emission frequency
\begin{equation}\label{Eq9}
\hbar\omega^{\text{quantum}}=M_n-M_{n-1}=\gamma\cdot{{\hbar}\over{32\pi
M}}\
\end{equation}
which is associated with the quantum transition $n\to n-1$ of the
black-hole area state in the asymptotic semi-classical $n\gg1$
regime.

I have therefore conjectured, in the spirit of the Bohr
correspondence principle \cite{Bek2,Born,Hod2}, that the
characteristic classical resonant frequency \cite{Notewaq} of these
spherically symmetric black holes should have the simple functional
relation [see Eqs. (\ref{Eq8}) and (\ref{Eq9})]
\begin{equation}\label{Eq10}
\omega^{\text{classical}}_{\text{R}}={{\ln k}\over{8\pi M}}\  ,
\end{equation}
where $k$ is the (yet unknown) integer parameter of Bekenstein and
Mukhanov \cite{Muk,BekMuk}.

The characteristic asymptotic resonant oscillation frequency of the
Schwarzschild black hole was, at that time, known only numerically
\cite{Noll}. In particular, using sophisticated numerical
techniques, it was computed with an accuracy of seven digits after
the decimal point \cite{Noll}:
\begin{equation}\label{Eq11}
M\omega^{\text{classical}}_{\text{R}}(n)=0.0437123\ \ \ \
\text{for}\ \ \ \ n\gg1\  .
\end{equation}

It was Saturday morning, twenty years ago \cite{Hodbr}, that I took
my calculator and, with a shaky hand, calculated the numerical
value of the expression $\ln2/8\pi$ [see Eq. (\ref{Eq10})].
To my deep disappointment, the answer I got did not agree with (\ref{Eq11}). But I did not lose
heart; I turned to calculate the value of $\ln3/8\pi$...and, to my
excitement, the calculator returned the answer $0.0437123$. It now
agreed, digit-by-digit, with the numerically computed \cite{Noll} characteristic
black-hole resonant oscillation frequency (\ref{Eq11}).

As a young student, it was amazing to feel, for the first time in my
(then) short scientific career, that I can count on my physical
intuition. Moreover, I was thrilled to know that I have found the
last missing piece in the black-hole quantization scheme of
Bekenstein and Mukhanov. In particular, substituting $k=3$ into Eqs.
(\ref{Eq1}) and (\ref{Eq8}), one obtains the discrete Bekenstein
area spectrum
\begin{equation}\label{Eq12}
A_n=4\hbar\ln3\cdot n\ \ \ ; \ \ \ n=1,2,3,...\
\end{equation}
of quantum Schwarzschild black holes.

It is worth emphasizing that the evenly spaced quantum black-hole
area spectrum (\ref{Eq12}), as originally formulated by Bekenstein
\cite{Bek1,Bek2} [see Eq. (\ref{Eq1})], is consistent both with the
thermodynamic entropy-area relation (\ref{Eq6}) of black holes, with the
Boltzmann-Einstein relation (\ref{Eq7}) of statistical physics, and,
as I have explicitly shown in \cite{Hod2}, with the Bohr correspondence
principle of quantum physics.

\section{Summary}

Jacob was known for his original and highly insightful physical
ideas, but not less for the beautiful and elegant ways he was using
to present them in his lectures and research papers. I, as many
others, benefited a lot from reading Jacob's seminal works on the
physics of black holes. His ingenious insights have inspired the
imagination of many physicists, I among them. Jacob's pioneering
ideas about the integration of quantum effects into black-hole
physics continue to inspire the physics community in her ongoing
search for a unified and self-consistent quantum theory of gravity
and thermodynamics.

As a final personal note, I would like to comment that,
unfortunately, I have never asked Jacob what he thought about my
idea to use the Bohr correspondence principle and the asymptotic
resonant frequencies of black holes in order to determine the value
of the dimensionless physical parameter $\gamma$ in his black-hole
quantum area spectrum (\ref{Eq1}). Now, that it is
too late, I can only regret for not doing so.

However, as fate would have it, many years after the appearance of
my paper \cite{Hod2} with the suggested black-hole quantization
spectrum (\ref{Eq12}), I came across a nice photo of Prof.
Bekenstein that was taken in his office in the occasion of him
winning the 2012 Wolf Prize in physics \cite{Bekln3}. In the
background of that photo I saw Jacob standing near his famous
blackboard, on which I could vividly see the fundamental factor of
$4\ln3$ \cite{Bekln3,Bekln3also}...

\bigskip
\noindent {\bf ACKNOWLEDGMENTS}

This research is supported by the Carmel Science Foundation. I would
also like to thank Yael Oren, Arbel M. Ongo, Ayelet B. Lata, and
Alona B. Tea for stimulating discussions.


\begin{thebibliography}{99}
\bibitem{Notetp} That is, for the classical absorption process to be an adiabatic one,
the absorbed particle should be characterized by a zero radial momentum.
\bibitem{Noteunit} We shall use natural units in which $G=c=1$. In these units $\hbar$
has the dimensions of length$^2$.
\bibitem{NoteBek2} As emphasized by Bekenstein \cite{Bek2}, the
Heisenberg uncertainty principle implies that, within the framework
of semi-classical general relativity, a quantized (finite-size)
particle cannot be absorbed precisely at the black-hole horizon from
an exact radial turning point of its motion.
\bibitem{NoteAvi} I would like to thank Dr. Avi Mayo for bringing
Jacob's seminal works \cite{Bek1} on the quantization of black holes
to my attention.
\bibitem{Notewaq} It is worth noting that black holes are
characterized by an infinite set $\{\omega_n\}^{n=\infty}_{n=1}$ of
damped gravitational quasinormal resonant modes. It is well known
that, within the framework of classical general relativity, black
holes do not emit radiation. Thus, in applying the Bohr
correspondence principle to the classical oscillation frequencies of
black holes, I have suggested \cite{Hod1,Hod2} to focus on the
highly damped ($n\gg1$) black-hole resonant modes, which are characterized by
infinitely small lifetimes (the characteristic zero lifetimes of
these highly damped modes correspond to the no emission status of
classical black-hole spacetimes).
\end{thebibliography}
\end{document}